\documentstyle[twocolumn,aps,psfig]{revtex}
\def\Journal#1#2#3#4{{#1} {\bf #2}, #3 (#4)}
\input{psfig}

\def\NPA{{\rm Nucl. Phys.} A}
\def\NPB{{\rm Nucl. Phys.} B}
\def\PLB{{\rm Phys. Lett.}  B}
\def\PRL{\rm Phys. Rev. Lett.}
\def\PRC{{\rm Phys. Rev.} C} 
\def\PRD{{\rm Phys. Rev.} D}
\def\PR{\rm Phys. Rev.}

\def\RMP{\rm Rev. Mod. Phys.}
\def\la{\langle}
\def\ra{\rangle}
\def\be{\begin{equation}}
\def\ee{\end{equation}}
\def\bea{\begin{eqnarray}}
\def\eea{\end{eqnarray}}
\def\lsim{\mathrel{\rlap{\lower4pt\hbox{\hskip1pt$\sim$}}
    \raise1pt\hbox{$<$}}}         
\def\gsim{\mathrel{\rlap{\lower4pt\hbox{\hskip1pt$\sim$}}
    \raise1pt\hbox{$>$}}}

\begin{document}
\tighten
\draft
\title{Pion Form Factor and Quark Mass Evolution in a Light-Front 
Bethe-Salpeter Model
}
\author{ L. S. Kisslinger$^a$, Ho-Meoyng Choi$^a$ and Chueng-Ryong Ji$^b$}
\address{$^a$Department of Physics, Carnegie-Mellon University,
Pittsburgh, PA 15213}
\address{$^b$Department of Physics, North Carolina State University,
Raleigh, NC 27695-8202}
\date{\today}
\maketitle
\narrowtext
\vspace{1.0in}
\begin{abstract}
We discuss the soft contribution to the elastic pion form factor
with the mass evolution
from current to constituent quark being taken into account using a
light-front Bethe-Salpeter (LFBS) model, which is a light-front quark
model (LFQM) with a running mass. 
It is shown that partial conservation of the axial-vector current (PCAC) is 
satisfied with a running quark mass.  We examine the sensitivity of 
the pion form factor using two different functional
forms of the quark propagator. The Ball-Chiu ansatz is used to maintain 
local gauge invariance of the quark-photon vertex. 
The extension of our model to the hard contribution is also discussed.
\end{abstract}
\ifpreprintsty
\vfill
\noindent
\hfill
\else
\fi
\narrowtext
\section{Introduction}

 The pion electromagnetic (EM) form factor is of great interest for the
study of Quantum Chromodynamics (QCD). At low momentum transfers ($Q^2$)
nonperturbative QCD (NPQCD) dominates, while at large $Q^2$ perturbative
QCD (PQCD) can be used to calculate the asymptotic form factor; and the
transition from NPQCD to PQCD has long been of
interest~\cite{BJPRSJR}. Since PQCD can
only be used for $Q^2 >$ 1 GeV$^2$, light-front (LF) quantization
methods may be most useful~\cite{BPP}.  
In the early work on the pion form factor\cite{JK} using the
light-front Bethe-Salpeter equation (LFBSE)\cite{LB} the NPQCD (soft)
part was separated from the PQCD (hard) part; and it was shown in a
model calculation that the transition from NPQCD to PQCD is expected in
the region 5.0 $ < Q^2 < 15.0$ GeV$^2$.
This work was extended\cite{KW}
to include the Sudakov form factor~\cite{SUD}, anomalous quark magnetic
dipole moments, and a simple model for the running quark mass, $m(Q^2)$.
Improved calculations of the pion EM form factor are
motivated by the recent new and upgraded data (up to $Q^2$=1.6 GeV) from
the Jefferson Laboratory(JLab)~\cite{JLab}.
While this $Q^2$ range may be
too low to determine the transition to the PQCD region,
these data are useful for studying NPQCD theoretical approaches.
In the present work we restrict ourselves to the soft NPQCD
part with models for the running mass that ensure gauge invariance and
consistency with the partially conserved axial current (PCAC) 
relation~\cite{GOR}
\be\label{PCAC}
m^2_\pi f_\pi=-2m_{(\nu)}\la 0|{\bar q}\gamma_5 q|\pi\ra_{(\nu)}=
-2m_{(\nu)}\frac{\la {\bar q}q\ra_{(\nu)} }{f_\pi},
\ee
where $m_{(\nu)}$=$m_0$(current mass) in spacelike
$p^2$=$\nu^2$ region($\nu$ is the renormlization point) and
$\la {\bar q}q\ra_{(\nu)}$ is the quark condensate.

 Since we are only considering the soft part of the pion form factor here,
the LFBS amplitude can be modeled by a light-front wave function~\cite{Dirac}
based on LF Hamiltonian dynamics, such as the 
light-front constituent quark model(LFCQM)~\cite{CQM1,CQM2,CJ}, but an
essential ingredient is the use of a running quark mass, which is the
main subject of the present paper.
In LF quantization, a possible connection is
anticipated between the constituent quark model(CQM)
and QCD due to
the rational energy-momentum dispersion relation that leads to a
relatively simple vacuum structure. There is no spontaneous creation of
massive fermions in the LF quantized vacuum. Thus, one can immediately
obtain a constituent-type picture, in which all partons in a hadronic
state are connected directly to the hadron instead of being simply
disconnected excitations (or vacuum fluctuations) in a complicated medium.
In particular, a systematic program has been laid out in the
calculation of the spacelike EM form factor of pseudoscalar mesons
because only parton-number-conserving Fock state (valence) contribution is
needed when the ``good" components of the current, $J^+(=J^0+J^3)$ and
$J_\perp=(J_x,J_y)$, are used in the
Drell-Yan-West($q^+$=0) frame~\cite{DY,LB}.
The new data from the JLab~\cite{JLab} seem to be in good
agreement with the previous CQM result~\cite{CJ} based on
a QCD-motivated linear confining potential. 
However, a possible realization of chiral symmetry breaking in the
LF vacuum is an underdeveloped aspect of LF quantization.

In contrast to quark models or LFCQM  which use a phenomenological
constant constituent quark mass, an approach based on QCD quantum field 
theory is the Bethe-Salpeter (BS) equation in conjuncton with Dyson-Schwinger
(DS) equations for the quark propagators, gluon propagator and vertices.
We note an important result of recent DS calculations, in which the 
effective running mass, $m(p^2)$, is calculated\cite{DS}. 
In these DS calculations
the parameters for model gluon propagators are fixed by fitting the
quark condensate, the mixed condensate and even the form of the
nonlocal condensate\cite{KM}. 
It is particularly interesting to note that the DS
quark propagator with running masses that decrease with respect to
$p^2$ faster than quark models gives properties of the rho\cite{KM1} 
and pion\cite{MT} that are in agreement with experiment. 
In all of these DS calculations it has
been found that the effective quark mass drops very rapidly with increasing
$Q^2$. 
Since the covariant BS/DS approach and the LF approach are not same but
complementary, it may be necessary to examine if the previous LFCQM 
result~\cite{CJ} is intact even if the quark mass evolves as rapid as
BS/DS approach found.
This is a strong motivation for
reformulating the LFCQM with a running quark mass, which we do in the
present work.

In the present work, we analyze the effect of the mass evolution (from 
constituent to current quark mass) on the elastic pion form factor at
low and intermediate $Q^2$. A correct theoretical approach to such a study
is the LFBSE coupled to LFDS equatons. From the LFDS equations the running
quark mass is obtained from the dressed quark propagator.
Although the LFDS equation has recently been 
developed~\cite{KL}, only simple model solutions are available, and here
we use a parameterization of
the running quark mass that is consistent with known observations.
While the asymptotic behavior of the running mass
might require the crossing symmetry(under $Q^2\leftrightarrow -Q^2$) at
high momentum $Q^2$ analogous to that of pion form factor, there is no
clue yet for the small momentum behavior in timelike region. Thus,
we present the two different forms of mass evolution
function; one is crossing asymmetric and the other is crossing symmetric.
We then compare the results for the two cases. It should be noted that the
recent LFDS results\cite{KL} also show a rapid decrease of effective
quark mass with momentum that puts the use of quark models for calculating
any but static properties in question. In the present work we use forms
for the quark mass evolution that are consistent with conventional quark
models for calculating hadronic properties at momentum transfers less than
about 800 MeV.

The paper is organized as follows: In Sec. II, we review the formulation
that underlies a description of the elastic pion form factor within
a modeling of QCD through the LFBSE, which is a LFQM with a running mass.  
In Sec. III, we formulate the running quark mass in LF framework and 
discuss the local gauge invariance at the quark-photon
vertex, i.e.  Ward-Takahashi identity~\cite{WT}, due to the momentum
dependent quark propagator. In Sec. IV, we analyze the running mass effect
on the pion form factor, charge radius, and decay constant obtained from
the previous CQM~\cite{CJ} calculation.
We also show that our model with the running quark mass is 
consistent with the PCAC relation given by Eq.~(\ref{PCAC}). This simply
means that we obtain a quark condensate consistent with the phenomenological
value given in Eq.~(\ref{PCAC}), which is true in the DS 
calculations\cite{DS,KM}.
A conclusion follows in Sec. V.
\section{Model Description}

  The LFBSE introduced in Refs.~\cite{JK,KW} has the form
\bea\label{LFBS}
\Psi(x,{\bf k}_{\perp})&=&\int [dy][d^2{\bf l}_\perp]
\biggl[{\cal K}_{\rm c}(x,{\bf k}_\perp;y,{\bf l}_\perp)\nonumber\\
& &\;\;\;\;\;
+ {\cal K}_{\rm g}(x,{\bf k}_\perp;y,{\bf l}_\perp)\biggr]
\Psi(y,{\bf l}_\perp),
\eea
where ${\cal K}_{\rm g}$ is the gluon exchange kernel and ${\cal K}_{\rm c}$,
the difference between the complete BS kernel and  ${\cal K}_{\rm g}$,
includes all confining effects. These kernels were obtained from the
relativistic string and LF perturbative QCD (PQCD), respectively.
Since NPQCD has not yet provided a form for the confining kernel,
in Ref.~\cite{KW} the problem of solving for the complete BS amplitude,
$\Psi(x,{\bf k}_{\perp})$, was avoided by using a model for the
soft amplitude. I.e., the soft BS amplitude,
$\Psi^{s}(x,{\bf k}_{\perp})$, can be considered to be the solution of
the equation
\be\label{BS_SF}
\Psi^{s}(x,{\bf k}_{\perp})\equiv
\int [dy][d^2{\bf l}_\perp]
{\cal K}_{\rm c}(x,{\bf k}_\perp;y,{\bf l}_\perp)
\Psi^s(y,{\bf l}_\perp).
\ee
Iterating Eq.~(\ref{LFBS}) by inserting $\Psi^{s}$ for $\Psi$, one obtains
the approximate form
\bea\label{LFBS2}
\Psi(x,{\bf k}_{\perp})&\approx& \Psi^{s}(x,{\bf k}_{\perp})\nonumber\\
&&+ \int [dy][ d^2{\bf l}_\perp]
{\cal K}_{\rm g}(x,{\bf k}_\perp;y,{\bf l}_\perp) \Psi^s(y,{\bf l}_\perp).
\eea
This BS amplitude contains both soft
and hard ingredients needed to take care of momentum transfer for all
$Q^2$ therefore is correctly characterized as including both confinement
and asymptotic features of a composite quark system.
This approach to the pion form factor has been
shown~\cite{KW} to be in good agreement with the direct BS
calculation~\cite{JK}, and to converge rapidly.
The extension of Eq.~(\ref{BS_SF}) to the non-wave-function vertex in the
particle-number-nonconserving Fock state contribution has recently been
discussed in Ref.~\cite{JC}.

 One nice feature of this approach is that one
determines the soft part and the hard part separately, so that one can
determine the transition from soft to hard QCD within this LFBS approach.
This will be the subject of our future work. 
For the present work of low- and medium-$Q^2$, we only
consider the confining part of the BS amplitude. We thus may be able
to use the LFCQM, which has inlcuded many important properties of the
$Q^2$ range that we focus in this work.
Therefore, in the rest of this work we use the LF wave function for the LFBS
amplitude.

\begin{figure}[t]
\centerline{\psfig{figure=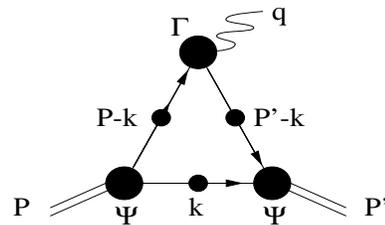,height=3cm,width=5cm}}
\caption{The valence diagram with dressed quark propagators
for the pion EM form factor calculation in $q^+=0$ frame 
where $p_q=P-k$, $p'_q=p_q-q=P'-k$ and $p_{\bar q}=-k$.
\label{pidress}}
\end{figure}

The elastic pion form factor is related to pion EM current
by the following equation
\be\label{Current}
\la P'|J^\mu(0)|P\ra = (P' + P)^\mu F_\pi(Q^2).
\ee
As usual, our calculation will be carried out using the $q^+=0$ frame where
$q^2=(P-P')^2=q^+q^--{\bf q}^2_\perp=-Q^2$, i.e. $Q^2>0$ is spacelike 
momentum transfer.

The matrix element of the current
given by Eq.~(\ref{Current}) can be expressed as a convolution integral
in terms of LF wave function, 
$\Psi^s(x,{\bf k}_\perp)$ as shown in Fig.~\ref{pidress}:
\bea\label{jmu}
\la P'|J^\mu(0)|P\ra &=&\sum_{\lambda_q\lambda'_q\lambda_{\bar q}}
\int^1_0 dx\int d^2{\bf k}_\perp
\Psi^{s*}_{\lambda'_q\lambda{\bar q}}(x,{\bf k'}_\perp)\nonumber\\
&\times&\frac{{\bar u}_{\lambda'_q}(p'_q)}{\sqrt{p'^+_q}}\Gamma^\mu
\frac{u_{\lambda_q}(p_q)}{\sqrt{p^+_q}}
\Psi^s_{\lambda_q\lambda{\bar q}}(x,{\bf k}_\perp),
\eea
where $p^+_q$=$p'^+_q$=$(1-x)P^+$ and 
${\bf k'}_\perp$=${\bf k}_\perp -x{\bf q}_\perp$ in the initial pion
rest frame, ${\bf P}_\perp$=0. The helicity of the quark(antiquark) is 
denoted as $\lambda_{q({\bar q})}$. 
Since the matrix element of the current
in Eq.~(\ref{jmu}) is symmetric under the exchange of $q$ and ${\bar q}$,
we do not explicitly write the contribution of
photon-antiquark interaction diagram as well as the charge factor.

Our LF wave function $\Psi^s$ in Eq.~(\ref{jmu}) is given by
\be\label{LFWF}
\Psi^s_{\lambda_q\lambda_{\bar q}}(x,{\bf k}_\perp)
=\sqrt{\frac{\partial k_z}{\partial x}}\Phi(x,{\bf k}_\perp)
{\cal R}_{\lambda_q\lambda_{\bar q}}(x,{\bf k}_\perp),
\ee
where $\Phi$ and ${\cal R}$ are the 
radial and relativistic spin-orbit wave functions, respectively. 
Our radial wave function is given by the gaussian trial function 
for the variational principle to the QCD-motivated effective LF
Hamiltonian\cite{CJ}:
\be\label{Rad}
\Phi({\bf k}^2)=(\frac{1}{\pi^{3/2}\beta^3})^{1/2}               
\exp(-{\bf k}^2/2\beta^2),
\ee
where ${\bf k}$=$(k_z,{\bf k}_\perp)$ is three vector and $\Phi({\bf k}^2)$
satisfies $\int d^3k|\Phi({\bf k}^2)|^2$= 1.
The LF variable ($x,{\bf k}_\perp$) is introduced in Eq.~(\ref{Rad})
by the definition of the longitudinal momentum $k_z$ via
$k_z$=$(x-1/2)M_0$ with $M^2_0$=$(m^2 + {\bf k}^2_\perp)/x(1-x)$.
If the quark mass depends on $x$ and ${\bf k}_\perp$, the Jacobian of the 
variable transformation
$(k_z,{\bf k}_\perp)$$\to$$(x,{\bf k}_\perp)$ in Eq.~(\ref{LFWF})
is obtained as 
\be\label{Jacob}
\frac{\partial k_z}{\partial x}=\frac{M_0}{4x(1-x)}
+ \frac{(2x-1)m(x,{\bf k}_\perp)}{2x(1-x)M_0}
\frac{\partial m(x,{\bf k}_\perp)}{\partial x}.
\ee
The spin-orbit wave function for a pseudoscalar meson
($J^{PC}=0^{-+}$) is obtained~\cite{CQM1,CJ}
 by the interaction independent Melosh transformation as follows:
\be\label{SO}
{\cal R}_{\lambda_q\lambda_{\bar q}}(p_q,p_{\bar q})=\frac{
{\bar u}(p_q,\lambda_q)\gamma_5 v(p_{\bar q},\lambda_{\bar q})}
{\sqrt{2}M_0}.
\ee

\section{Quark Mass Evolution and Local Gauge Invariance}
\begin{figure}
\centerline{\psfig{figure=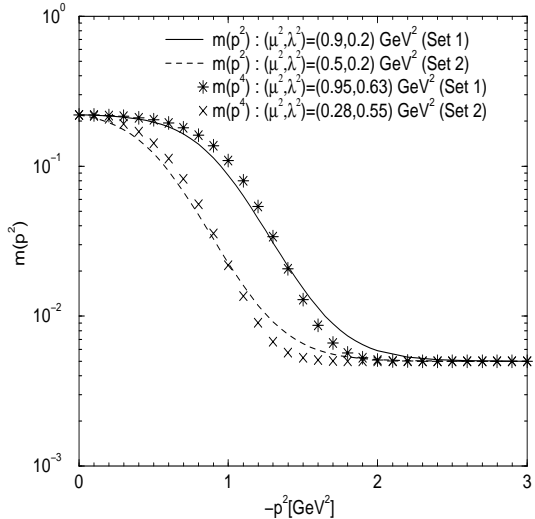,height=80mm,width=80mm}}
\caption{Quark mass evolution in spacelike momentum region,
$-p^2>0$.\label{massfig}}
\end{figure}

The solution of the DSE for the renormalized dressed-quark 
propagator takes the form in Minkowski space
\be\label{QP}
S(p)^{-1}=A(p^2){\not\! p} - B(p^2),
\ee
where the quark mass evolution function $m(p^2)$ is defined as 
$m(p^2)=B(p^2)/A(p^2)$. Also,
the gauge invariance requires that the quark-photon 
vertex $\Gamma^\mu$ given by Eq.~(\ref{jmu}) satisfy the vector
Ward-Takahashi identity(WTI)~\cite{WT}(i.e. current conservation)
\be\label{WTI}
-q^\mu\Gamma_\mu(p;q)= S(p')^{-1}-S(p)^{-1},
\ee
where $q=p - p'$. At zero momentum transfer $q$=0, the quark-photon vertex
is also specified by the differential Ward identity (i.e. charge conservation)
$\Gamma^\mu(p;0)=\partial S(p)^{-1}/\partial p_\mu$.
The bare quark-photon vertex, $\Gamma^\mu=\gamma^\mu$, which is usually used
in LFCQM~\cite{CQM1,CQM2,CJ}, is inadequate when the quark propagator 
has momentum-dependent dressing because it violates WTI.
As used in many DSE studies of EM interactions~\cite{MRT,DS}, 
we take the Ball-Chiu(BC) ansatz~\cite{BC} for 
the quark-photon vertex 
\bea\label{BC}
\Gamma^\mu_{\rm BC}&=& \frac{ ({\not\! p}
+ {\not\! p'})}{2}(p + p')^\mu
\frac{ A(p'^2)-A(p^2)}{p'^2-p^2} \nonumber\\
&+& \frac{A(p'^2)+A(p^2)}{2}\gamma^\mu
- (p + p')^\mu\frac{B(p'^2)-B(p^2)}{p'^2-p^2}.
\eea
Here, we introduce two algebraic parametrizations of the quark running 
mass, i.e. crossing asymmetric(CA) and crossing symmetric(CS) 
mass functions proportional to $p^2$ and $p^4$, respectively. 
For the CA mass evolution function, we take 
\be\label{ME} 
m(p^2)=m_0 + (m_c - m_0)\frac{ 1+\exp(-\mu^2/\lambda^2)}
{1 +\exp[(-p^2 - \mu^2)/\lambda^2]},
\ee
where $m_0$ and $m_c$ are the current and constituent quark masses, 
respectively. The parameters $\mu$ and $\lambda$ are used to adjust the
shape of the mass evolution. Similarly, the following form of the
mass evolution function is used for the CS case 
\be\label{ME2}
m(p^4)=m_0 + (m_c - m_0)\frac{ 1+\exp(-\mu^4/\lambda^4)}
{1 +\exp[(p^4 - \mu^4)/\lambda^4]},
\ee
where we simply replace $-p^2$ and $\mu^2(\lambda^2)$ in Eq.~(\ref{ME})
with $p^4$ and $\mu^4(\lambda^4)$, respectively. 
In our case, we set $A(p^2)$=1, $B(p^2)$=$m(p^2)$ for CA
and  $B(p^2)$=$m(p^4)$ for the CS case, respectively.
While Eqs.~(\ref{ME}) and~(\ref{ME2}) are phenomenological, 
the results of our running mass in spacelike momentum region
($-p^2>0$) yield a generic picture
of the quark mass evolution from the low energy 
limit of the constituent quark mass to the high energy 
limit of the current quark mass. 
For comparison, we use in Fig.~\ref{massfig} two different parameter 
sets for each mass evolution function, i.e., 
$(\mu^2,\lambda^2)=(0.9,0.2)$ [Set 1] and $(0.5,0.2)$ [Set 2]
(in unit of GeV$^2$) for $m(p^2)$ and $(\mu^2,\lambda^2)=(0.95,0.63)$ [Set 1]
and $(0.28,0.55)$ [Set 2] (in unit of GeV$^2$) for $m(p^4)$, 
respectively (See Fig.~\ref{massfig} for the line code). 
The current and constituent quark 
masses used are $m_0=5$ MeV and $m_c=220$ MeV, respectively. 
Simulating the constituent picture at small momentum region, 
we have chosen these particular sets of parameters, [Set 1] and [Set 2] 
for each mass function, to keep the constituent mass up 
to $(-p^2)\sim 1$ and 0.5 GeV$^2$, respectively, before it drops 
exponentially.

In order to express the four momentum $p^2$ in terms of LF variables
$(x,{\bf k}_\perp)$, we use the on-mass shell condition, $p^2=m^2(p^2)$.
It implies zero binding energy of a mock meson, i.e. 
$P^-=p^-_q + p^-_{\bar q}$ where $P^-(=P^0-P^3)$ and $p^-_q(p^-_{\bar q})$ 
are the LF energies of the mock meson and the quark(antiquark),
respectively. It leads to the following identity for the pion case
($m_q=m_{\bar q}$)
\be\label{ZeroBE}
p^2 = x(1-x){\tilde M}^2 - {\bf k}^2_\perp.
\ee
For the mock meson mass ${\tilde M}$, we
take the average value(so called spin-averaged meson mass) of 
$\pi(m_\pi)$ and $\rho(m_\rho)$ masses with appropriate weighting factors 
from the spin degrees of freedom, i.e. ${\tilde M}$=$(m_\pi 
+ 3m_\rho)_{\rm exp}/4$=612 MeV, which is consistent with quark model
calculations with typical constituent quark masses and been used in some 
constituent LFQM calculations~\cite{CQM2}. Note that
the spacelike form factor $F_\pi(Q^2)$ is weakly 
dependent on ${\tilde M}$, i.e. the numerical result with the mock
meson mass ${\tilde M}=612$ MeV is not much different from that with 
the physical pion mass $m_\pi=140$ MeV, which would be used in the
LFBS-LFDS approach.
Using Eq.~(\ref{ZeroBE}), we can now express 
the mass evolution functions $m(p^2)$ and $m(p^4)$ in terms 
of LF variables $x$ and ${\bf k}_\perp$.

  Using the good component of the current, $J^+$,
the pion EM form factor in Eq.~(\ref{Current}) is obtained as 
\bea\label{jp}
F_\pi(Q^2) 
&=&N_{\pi}\int dx d^2{\bf k}_\perp
\sqrt{\frac{\partial k'_z}{\partial x}}\sqrt{\frac{\partial k_z}{\partial x}}
\Phi^*_f(x,{\bf k'}_\perp)\Phi_i(x,{\bf k}_\perp)
\nonumber\\
&\times&\biggl\{ 
\frac{{\bf k}_\perp\cdot{\bf k'}_\perp + m_k m_{k'}}
{x(1-x)M_0M'_0} +
\frac{m_k{\Delta m_{k}}(2{\tilde M}^2 + {\bf q}^2_\perp)}
{M_0M'_0{\Delta{\bf k}^2_\perp}}\biggr\},\nonumber\\
\eea
where $N_\pi$ is the normalization constant and 
$\Delta m_{k}=m(x,{\bf k'}_\perp)-m(x,{\bf k}_\perp)$=$m_{k'}-m_k$ and 
$\Delta{\bf k}^2_\perp={\bf k'}^2_\perp-{\bf k}^2_\perp$.
The other primed terms, $k'_z$ and $M'_0$, are obtained from 
$k_z(x,{\bf k}_\perp\to{\bf k'}_\perp)$ and
$M_0(x,{\bf k}_\perp\to{\bf k'}_\perp)$, respectively.
The terms in the curly bracket are obtained from the
trace of spin-orbit wave function, i.e. 
$\sum{\cal R}^\dagger({\bar u}/\sqrt{p'^+_q})\Gamma^+_{\rm BC}
(u/\sqrt{p^+_q}){\cal R}$.
Note that the normalization constant
$N_\pi$ at $Q^2=0$ is exactly one in chiral limit ($m_0={\tilde M}=0$).  

We also obtain the quark condensate from the PCAC relation given
by Eq.~(\ref{PCAC}) as follows:
\bea\label{qq}
\la{\bar q}q\ra&=&-\frac{f_\pi\sqrt{N_c}}{(2\pi)^{3/2}}
\int\frac{dx d^2{\bf k}_\perp}{x(1-x)}
\sqrt{\frac{\partial k_z}{\partial x}}\Phi(x,{\bf k}_\perp)
\sqrt{m^2_k + {\bf k}^2_\perp},\nonumber\\
\eea
where $N_c$=3 is the color factor. This is equivalent to the expression
in DS models\cite{DS} for A(p$^2$) = 1. The quark condensate is normally
evaluated at the spacelike momentum scale $p\sim 1$ 
GeV(corresponding to the renormalization point $\nu$~\cite{DS})
where $m_k\to m_0$(see Fig.~\ref{massfig}). 

\section{Numerical Results}

In our numerical calculations, we use the model parameters 
$(m_c,\beta)$=(0.22,0.3659) [GeV] obtained in Ref.~\cite{CJ} for the linear
confining potential model where the 
charge radius($r^2_\pi$=$-6dF_\pi(Q^2)/dQ^2|_{Q^2=0}$) and 
decay constant($\la 0|{\bar q}\gamma^\mu\gamma_5 q|P\ra=if_\pi P^\mu$) 
of the pion were predicted as $r^2_\pi$=0.425 
[fm$^2$](Exp. = 0.432$\pm$0.016~\cite{Amen}) 
and $f_\pi$=130 MeV (Exp. = 131 MeV~\cite{PDG}), respectively. 
The change of the charge radius and decay constant from the CQM 
result due to the running mass formulae are within 2$\%$ and 5$\%$, 
respectively. For the calculation of the quark condensate, 
we obtain, for example, $f_\pi$= 127 MeV from the [Set 1] of the CA
mass function. Consequently, we obtain from
Eqs.~(\ref{PCAC}) and~(\ref{qq}) the pion mass and the quark 
condensate as $m_\pi$=164 MeV and $-\la{\bar q}q\ra$=(0.3 GeV)$^3$
while the experimental values of $m_\pi$ and $-\la{\bar q}q\ra$ are
140 MeV and (0.236 GeV)$^3$~\cite{qqE}, respectively. 
This shows the PCAC relation is reasonably well satisfied in LFQM 
with our mass 
evolution function. The results from other parameter sets are
not much different from the above [Set 1] of CA case. 

\begin{figure}
\centerline{\psfig{file=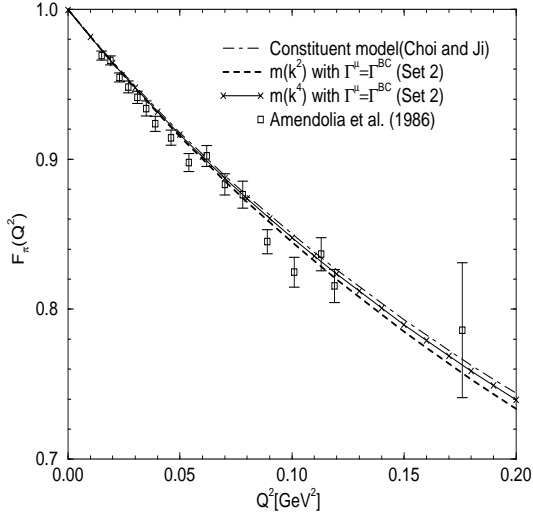,height=80mm,width=80mm}}
\caption{Pion EM form factor for small $Q^2$ region using the
[Set 2] for both CA and CS mass formulae compared with the
experimental data~\protect\cite{Amen} and the CQM result~\protect\cite{CJ}.
\label{PiForm}}
\end{figure}

\begin{figure}
\centerline{\psfig{file=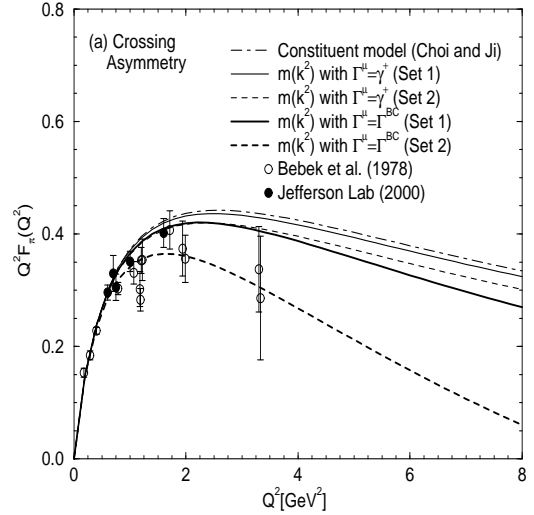,height=80mm,width=80mm}}
\vspace{0.1cm}
\centerline{\psfig{file=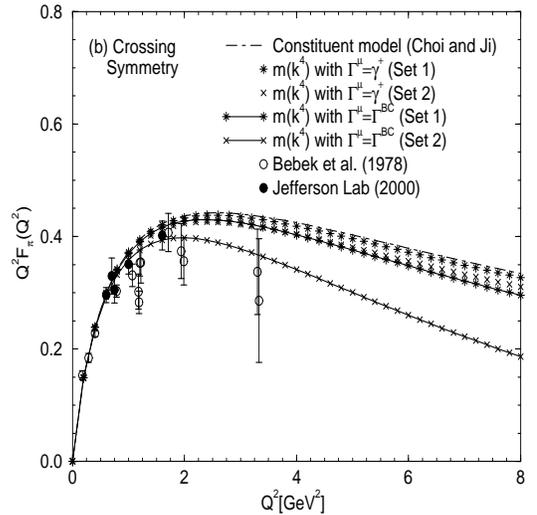,height=80mm,width=80mm} }
\caption{Pion EM form factor:(a) Crossing asymmetry(CA) and
(b) Crossing symmetry(CS) mass functions compared with the experimental
data~\protect\cite{JLab,Bebek} as well as the CQM result~\protect\cite{CJ}.
\label{pi1}}
\end{figure}

In Fig.~\ref{PiForm}, we show our results of the pion EM form factor
for small $Q^2$ region using the [Set 2] with BC vertex for both CA and
CS mass evolution functions and compare with the experimental data~\cite{Amen}
as well as the CQM result given by Ref.~\cite{CJ}. The line code is in the
figure. The small momentum($Q^2$) behavior of the form factor with CA
and CS mass functions are not only close to each other but also 
close to the experimental data~\cite{Amen} as well as the CQM 
result~\cite{CJ}. The results of [Set 1] for both CA and CS mass functions
are even closer to the CQM result.
We also show in Fig.~\ref{pi1} our results of the form factor 
for the intermediate $Q^2$ region for CA [Fig.~\ref{pi1}(a)]
and CS  [Fig.~\ref{pi1}(b)] mass functions compared with
the experimental data~\cite{JLab,Bebek} as well as the CQM result~\cite{CJ}. 
The line code is given in each figure.   
As one can see from Fig.~\ref{pi1}, (1) the difference between the
bare vertex and BC ansatz indicates the breakdown of the local gauge
invariance from the usage of the bare vertex, 
(2) the [Set 2] for both CA and CS mass functions show larger deviation 
from the CQM result than the [Set 1] case for the momentum transfer 
$Q^2\sim$2 GeV$^2$ and above region,
(3) the results with BC vertex fall off faster (at around $Q^2$=2 GeV$^2$) 
than the CQM result does, (4) the mass evolution effects from
[Set 1] for both CA and CS cases are not much different from the constituent 
result up to $Q^2$=8 GeV$^2$, and (5) the CA mass evolution function is more
sensitive to the variation of the momentum dependence than the CS mass 
evoluton function. 

\section{Conclusion}
In conclusion, we have reexamined the soft contribution to the
pion elastic form factor in the framework of the LFBS with
a $Q^2$-dependent quark mass that could be obtained from a LFDSE.
This is equivalent to the LFQM with a running quark mass.
The Ball-Chiu ansatz was used for the dressed  quark-photon vertex.
We showed that the PCAC relation in Eq.~(\ref{PCAC}) is satisfied in LFQM 
with our running mass formulae. 
The CQM result is not affected too much by the quark mass 
evolution for the small momentum transfer region up to 
$Q^2$=1$\sim$2 GeV$^2$. 
However, the form factor is sensitive to
the shape of the mass evolution for the intermediate ranges, e.g.
the [Set 1] for both CA and CS cases are not much different from the
LFQM result but the [Set 2] show a sizable effect
on the soft pion elastic form factor for $Q^2\sim$ 2 GeV$^2$ and above.
It may be interesting to observe from [Set 2] 
that the form factor may distinguish the mass 
evolution respecting CS from the one not respecting CS even though 
they show similar momentum dependent behavior as shown in Fig.~\ref{massfig}.
While our calculation employed a phenomenological mass evolution 
functions and the quark-photon vertex modification, 
the qualitative feature presented in this work might not be
significantly modified even if one were to use the more realistic
solutions obtained from the first principle. 
However, it may be interesting to check further   
whether the soft part would fall more rapidly
if the present results of DS\cite{DS} and LFDS\cite{KL} models were used.  
The ``hard-scattering" contribution to the pion form factor, i.e.
the second term in Eq.~(\ref{LFBS2}),
is under investigation.

\vspace{0.5 cm}
\centerline{\bf Acknowledgements}
\bigskip

  The authors would like to acknowledge helpful discussions with Otto
Linsuain and Pieter Maris. 
The work of L.S.K. and H-M.C was supported in part by the NSF
grant PHY-00070888 and that of C-R. Ji by the US DOE under grant
No. DE-FG02-96ER40947. The North Carolina Supercomputing Center and
the National Energy Research Scientific Computer Center are also
acknowledged for the grant of Cray time.

\end{document}